\documentclass[aps,preprint]{revtex4}
\usepackage{amsfonts}
\usepackage{amsmath}
\usepackage{amssymb}
\usepackage{graphicx}
\usepackage{epsfig}

\begin{document}
\preprint{LEZ/00104}
\title{Wave Functions of Heliumlike Systems in Limiting Regions}

\author{E. Z. Liverts}
\affiliation{Racah Institute of Physics, The Hebrew University,
Jerusalem 91904, Israel}
\author{M. Ya. Amusia}
\affiliation{Racah Institute of Physics, The Hebrew University,
Jerusalem 91904, Israel; A. F. Ioffe Physical-Technical Institute,
St. Petersburg, 194021, Russia}
\author{E. D. Drukarev}
\affiliation{Petersburg Nuclear Physics Institute,
Gatchina, St. Petersburg 188300, Russia}
\author{R. Krivec}
\affiliation{Department of Theoretical Physics, J. Stefan Institute,
P.O. Box 3000, 1001 Ljubljana, Slovenia}
\author{V. B. Mandelzweig}
\affiliation{Racah Institute of Physics, The Hebrew University,
Jerusalem 91904, Israel}

\pacs{31.15.Ja, 31.15.-p, 31.10.+z}

\begin{abstract}
We find approximate analytical presentation of the solutions
$\Psi(r_1, r_2, r_{12})$ of Schr\"odinger equation for two-electron system
bound by the nucleus, in the space region $r_{1,2}=0$ and $r_{12}=0$ that are
of great importance for a number of physical processes.
The presentation is based on the well known behavior
of  $\Psi(r_1, r_2, r_{12})$  near the singular
triple coalescence point. The approximate functions are compared to the locally correct
ones obtained earlier by the
Correlation Function Hyperspherical Harmonic (CFHH) method
for helium atom, light heliumlike ions and for the negative ion of hydrogen $H^-$.
The functions are shown to determine a natural basis for the
expansion of CFHH functions in the considered space region.
We demonstrate, how these approximate functions simplify the calculations
of the high energy ionization processes.
\end{abstract}


\maketitle

\section{Introduction}

The ground states of the two-electron systems bound by the nucleus are
described by the wave functions, depending on three variables. These can
be the distances between the electrons and the nucleus
$r_{1,2}$ and the interelectron distance $r_{12}$. Here we shall find
analytical expressions, which would approximate the solutions of the Schr\"odinger equation
$\Psi(r_1, r_2, r_{12})$ in the special cases $r_{1,2}=0$ and
$r_{12}=0$:
\begin{equation} \label{1}
F(R)\equiv \Psi(0,R,R); \quad
\Phi(R)\equiv \Psi(R,R,0)\ ,
\end{equation}
We shall consider the ground states of helium atom and of the light heliumlike ions,
including also the negative ion of hydrogen $H^-$.
In this paper we shall treat the ground states only. Thus, the total spin of the
two-electron system is equal to zero.

Note that this problem differs strongly from the traditional problem of approximating
the total wave function  $\Psi(r_1, r_2, r_{12})$ \cite{1}. There are numerous wave functions
of this kind, with the approximate functions being usually certain combinations of exponentials
and polynomials, while a set of fitting parameters is found by minimization of the energy
functional. Thus, the quality of such functions is determined by the accuracy of reprodusing
the binding energy value. Since the averaged value of the Hamiltonian is determined by the
distances, which are of the order of the size of the atom, such functions
provide very good approximation at these distances. However, as it was emphasized already in
\cite{2}, they are not necessary as precise in the limiting cases $r_{1,2}=0$ and $r_{12}=0$.

The motivation of our study is that in a number of dynamical problems one needs the bound
state wave functions in the such region of variables, where one of the distances is
much smaller then the other ones. This takes place in those
processes on the bound electrons, which are kinematically forbidden for the free ones.
For example, the high energy asymptotic of photoionization is expressed in terms
of the two-electron function $\Psi(0,R,R)$, where $r_{1}$ or $r_{2}$ is zero \cite{1}.
The same is correct for the nonrelativistic high energy asymptotic of the double
photoionization, and for the energy distribution
of the Compton scattering at sufficiently small energies of the outgoing electrons.
Some of the characteristics of the double photoionization are expressed in terms of the
two-electron function with zero interelectron distance $\Psi(R,R,0)$ \cite{3}.
The straitforward way to obtain the functions $F(R)$ and $\Phi(R)$ is to
calculate them from $\Psi(r_1, r_2, r_{12})$, that could be derived numerically.
This is a rather complex procedure.
Therefore here we shall build the approximate wave functions $F_A(R)$ and $\Phi_A(R)$ based
on the known behavior of the exact wave function only near the triple coalescence point $R=0$.
The only free parameter of our approach is the value of the wave functions at the coordinate origin
\begin{equation} \label{2}
N\ =\ \Psi(0,0,0)\ .
\end{equation}
Our approach was initiated and encouraged by the large role of the proper treatment of the two-particle
coalescence point in the earlier calculations.
For example, the binding energies can be reproduced usually with a good accuracy
by the approximate wave functions, which are certain combinations of
exponential and polynomial factors \cite{1}. Such presentation is good
enough at distances of the order
of the atomic size. However, it was understood long ago
that the analytical dependence  upon $R$ is not as simple as that,
and the logarithmic terms are presented in the expansion of
the wave function near the origin \cite{4}. Later it was
found that, if
$r_{1,2}$ or $r_{12}$ turn to zero,
the solution of the Schr\"odinger equation should satisfy the specific
Kato conditions \cite{5}.
Inclusion of the logarithmic terms \cite{6}, or accounting of
the Kato conditions \cite{7} or both \cite{8} does not
influence much the energy value, but improves strongly the convergence
of the procedure of $\Psi(r_1, r_2, r_{12})$ calculations.
This encouraged us to try a rather simple approach.

As it is shown in this paper, the approximate functions for (1) appear to be
\begin{equation} \label{3}
F_A(R)=N\exp\left[-\left(Z-\frac{1}{2}\right)R\right]; \quad
\Phi_A(R)=N\exp(-2ZR)\ .
\end{equation}

They have to be compared to precise or highly accurate locally correct functions
$F_{LC}(R)$ and $\Phi_{LC}(R)$. As such, we use the functions obtained by the Correlation
Function Hyperspherical Harmonic (CFHH) method \cite{9}.
These non-variational wave
functions of the two-electron system bound by a light nucleus in
$s$-state have been obtained by direct solution of the
three-body Schr\"odinger equation \cite{10}, without additional
approximations. They require complicated computer codes for being employed.

The way we construct the approximate wave functions insures that they reproduce
 the CFHH functions $F_{LC}(R)$ and $\Phi_{LC}(R)$ with good accuracy at sufficiently
small values $R$. The question is, how long this can last, while $R$  increases.
In other words, we must calculate the characteristics of the processes, which are determined
by $F(R)$ and $\Phi(R)$ at $R$ being of the order of the size of the
atom, and compare the results obtained with (3) and with the CFHH
functions.

The answer is that the relative discrepancy between functions (3) and CFHH functions
does not exceed several percents at characteristic distances
$\frac{1}{(Z-1/2)}$ and $\frac{1}{2Z}$. The same is the accuracy of the photoionization
characteristics.

Of course, such accuracy would not have been sufficient for the calculation of the static
atomic characteristics, e.g. of the energy levels values. However, e.g. there was
qualitative controversy in theoretical results on the double photoionization energy
distribution until recent time \cite{11}, with quantitative results differing by orders
of magnitude. Thus it would be unjustified to run for too high accuracy in any case.
On the other hand,
good accuracy of the functions (2) prompts a basis for expansion of the CFHH
functions. Since the functions (3) have the radial dependence of the $1s$-functions in the Coulomb
fields with charges $(Z-\frac{1}{2})$ and $2Z$, respectively, one can present the numerical
CFHH functions as linear combinations of the functions of this field with the dominative
contribution coming from $1s$ terms.

We build our approximate wave functions and discuss their relation to other approaches in Sec.2.
We analyze expansion of CFHH functions at two-particle coalescence points in series of
the single particle eigenfunctions of Coulomb fields in Sec.3.
We consider the applications in Sec.4, and summarize in Sec.5. Atomic system of units is used
through the paper.

\section{Wave functions}

It is known that at small distances $r_{1,2} \ll Z^{-1}$ the solution of the
Schr\"odinger equation can be presented as \cite{12,13}
\begin{equation} \label{4}
\Psi(r_1,r_2,r_{12})\ =\ N\left[1-
Z(r_1+r_2)+\frac12 r_{12}+O(r^2,r^2\ln r)\right],
\end{equation}
with $r=\sqrt{r_1^2+r_2^2}$. The explicit form of the quadratic terms
was found in \cite{12}. Eq.(4) is consistent with more general Kato conditions
\cite{5},
\begin{eqnarray} \label{5}
&&\left.\frac{\partial\Psi(r_1,r_2,r_{12})}{\partial
r_1}\right|_{r_{1}=0}\ =\ -Z\Psi(0,r_2,,r_2)\,;
\nonumber\\
\nonumber
&& \left.\frac{\partial\Psi(r_1,r_2,r_{12})}{\partial
r_2}\right|_{r_{2}=0}\ =\ -Z\Psi(r_1,0,r_1)\,;
\\
&& \left.\frac{\partial\Psi(r_1,r_2,r_{12})}{\partial
r_{12}}\right|_{r_{12}=0}=\ \frac12\,\Psi(r_1,r_1,0)\,.
\end{eqnarray}
 which are fulfilled for the CFHH functions. Using Eq.(4) we find
 that at  $r_{1,2} \ll Z^{-1}$
\begin{equation} \label{6}
F(R)=N\left[1-\left(Z-\frac12\right)R+...\right]; \quad
\Phi(R)=N\left(1-2ZR+...\right)\,,
\end{equation}
with the dots denoting the higher terms. This provides
\begin{equation} \label{7}
\lim\limits_{R\to0}\frac1{F(R)}\,\frac{dF(R)}{dR}\ =\ -Z+\frac12\ ,
\end{equation}
and
\begin{equation} \label{8}
\lim\limits_{R\to0}\frac1{\Phi(R)}\frac{d\Phi(R)}{dR}\ =\ -2Z\ .
\end{equation}

We require Eqs. (7) and (8) to be satisfied by
our approximate functions $F_A(R)$ and $\Phi_A(R)$ for all $R$.
This leads to Eq.(3).

The functions (3) correspond to a very simple physical pictures. Note that (3)
look like the $1s$ functions in the Coulomb fields with
charges $(Z-1/2)$ and $2Z$, respectively, that serve in fact as a sort
of adjustable parameters.
It's $R$ dependence is one of $1s$ electron,
while the small probability of the three-particle
coalescence is contained in the factor $N$ determined by Eq.(2).
We calculate the latter by using the CFHH functions.

To characterize the
quality of our approximate functions,
we introduce the
\begin{equation} \label{9}
y_1(R))=\log_{10}|\frac{F_A(R)-F_{CFHH}(R)}{F_{CFHH}(R)}|;\quad
y_2(R)=\log_{10}|\frac{\Phi_A(R)-\Phi_{CFHH}(R)}{\Phi_{CFHH}(R)}|\,
\end{equation}
with the lower indices CFHH denotes the wave functions obtained in
\cite{10}.

The accuracy of the functions (3) increases rapidly with the nuclear
charge $Z$ growth. However even for the negative ion $H^-$
($Z=1$) the accuracy is rather high. At characteristic $R \sim (Z-1/2)^{-1}$,
and $R \sim (2Z)^{-1}$
the errors of the function $\Phi_A$ for H$^-$ make 6\%, being less than
1\% for the function $F_A$. The errors increase at larger values of
$R$.  They exceed the value of 10\% at the distances, at which the wave
functions are already very small. The functions $y_i(R)$ (9),
describing $R$-dependence of the errors are presented in Fig. 1.
We present the results for helium ($Z=2$) since most of the studies of
the two-electron systems are carried out for this case. We give also the
results for $Z=4$ to illustrate $Z$ dependence. A curve
for $H^-$ ($Z=1$) is also presented, since this case is the most difficult for investigations.
The dip on the graph of Fig. 1a is a result of the logarithmic scale, since the logarithm
of the absolute value of the difference of the two functions goes to $-\infty$
at the points where the difference changes sign. The overall accuracy of the solution
therefore can be inferred only at the values of R not too close to the dip.

One can see that the discrepancy with CFHH functions  becomes much greater
at $R$ becoming of the order of the size of the atom, comparing to
that at smaller $R$. However, the precision is still good enough for obtaining
results with the accuracy of several percents.

The values of $N$, the latter being defined by Eq.(2) are presented in
Table 1. At large $Z$ the single-particle hydrogenlike model is
expected to become increasingly true, since the interaction between the
electrons is $Z$ times weaker than their interaction with the nucleus.
Hence, in the limit $Z\gg1$
\begin{equation} \label{10}
N\ =\ N_c\ =\
\frac{Z^3}\pi\ .
\end{equation}
The results presented in Table 1
illustrate this tendency. As expected, deviations from the limiting law
(10) are of the order $Z^{-1}$. The actual results are smaller than
predicted by (10) since the latter does not include the electron
repulsion, which diminishes this value.

Of course, there are numerous simple approximate wave functions of the type
$$ \Psi_A(r_1,r_2,r_{12})=c(exp(-ar_1-br_2)+exp(-ar_2-br_1)),$$
which are build in order to calculate the ground state energy values \cite{1},
thus approximating the solutions of the  Schr\"odinger equation at
$r_{1,2}$ of the order $Z^{-1}$ (in the case of $H^{-}$ they had to reproduce also the very
existence of the bound state). Technically, they turn to the single-exponential forms at
$r_{1}=r_{2}=R$ and do not depend on $r_{12}$. These functions can be compared
to our functions $\Phi(R)$ defined by Eq.(3).
But they do not approximate the locally correct CFHH
functions $\Phi_{A}(R)$, and, following \cite{2}, are not supposed to. We illustrate this statement
by presenting in Fig.2 the CFHH function $\Phi(R)$, our function (3) and the screened Coulomb wave
function $\Phi_s(R)=\frac{a^3}{\pi}exp(-2aR)$ with $a=27/16$ for helium \cite{1}.

In \cite{14}  the function $F(R)$ for $H^{-}$, $He$ and $Li^{+}$ was approximated by a hydrogenlike
function with the effective charge $Z_{eff}$ treated as a variational parameter.
The values of $Z_{eff}$ for $Z=1,2,3$ have been found to be 0.58, 1.53 and 2.52, correspondingly.
In \cite{15} the function $F(R)$ for the ion $H^{-}$ have been analyzed at large distances.
We do not claim our functions to be true in this $R$-region, which is not essential
for us since of prime importance is the $R$ domain within the atomic radius.

\section {Expansion of CFHH functions in series of the Coulomb field
eigenfunctions}

The $R$ dependence of the approximate wave functions $F_A(R)$ and
$\Phi_A(R)$ (3) is the same as that of $1s$ functions in the Coulomb
fields of the nuclei with the charges $Z_1=Z-1/2$ and $Z_2=2Z$, respectively.
The high precision of these functions suggests that the
eigenfunctions of the Schr\"odinger equations in these fields compose
convenient series for expansion of the CFHH functions $F(R)$ and
$\Phi(R)$.

Introducing the common notation $X(R)$ for the functions $F(R)$ and
$\Phi(R)$ we present the normalized functions
$X_N(R)=\frac1{C^{1/2}_X}X(R)$ with $C_X=\int\limits^\infty_0 R^2X^2(R)dR$. Thus $\int\limits^\infty_0
R^2X^2_N(R)dR=1$.

In the expansions over the complete sets of some
eigenfunctions, $X_{N}(R)$ can be presented as:
\begin{equation} \label{11}
F_N(R)=\sum a_if_i(R); \qquad \Phi_N(R)=\sum b_i\varphi_i(R)\ ,
\end{equation}
with $\sum$ denoting the sum over the states of discrete spectrum and
integration over continuum
\begin{equation} \label{12}
a_i=\int\limits^\infty_0 R^2F_N(R)f^*_i(R)dR; \qquad b_i=\int\limits^\infty_0
R^2\Phi_N(R)\varphi^*_i(R)dR\ .
\end{equation}
For $f_i(R)$ and $\varphi_i(R)$ normalized to one, it is
\begin{equation} \label{13}
\sum a^2_i\ =\ \sum b^2_i\ =\ 1\ .
\end{equation}
Choosing the solutions of the Schr\"odinger equations in the Coulomb
fields with the charges $Z_1=Z-1/2$ and $Z_2=2Z$ as the functions
$f_i(R)$ and $\varphi_i(R)$ respectively, we find the values
$a_{1s}$ and $b_{1s}$ presented in Table 2. For atomic helium
$a_{1s}=0.9997$, $b_{1s}=0.998$. High accuracy of the functions (3)
corresponds to domination of the terms $a^2_{1s}$ and $b^2_{1s}$ in the
sums (13).

The precision of calculations can be improved by adding the contributions of
the higher states according to Eq.(12). Of course, in our case only the
$s$-states are involved. For example, $a_{2s}=-0.02$, $b_{2s}=-0.05$ in
the case of atomic helium. The results for the other values of $Z$ are
presented in Table 2. This procedure enables to achieve any desired
accuracy, controlled by Eq. (13).

\section{Examples of application}

As we said above, one of the possible application of the functions (3) is
the high energy photoionization processes. Let us start with the single photoionization.
The high energy nonrelativistic asymptotic for the K-shell ionization cross section
can be written as \cite{1}
\begin{equation} \label{14}
\sigma =\frac{2^{11/2} \pi e^2 Z^2 C^2}{3 m c \omega^{7/2}}\ ,
\end{equation}
where $m$ is electron mass ans $c$ is the speed of light.
The properties of the ionized states contained in the factor
\begin{equation} \label{15}
C=\int\limits^\infty_0 R^2 F(R)\psi_K(R)dR\ .
\end{equation}
Here F(R) is determined by Eq.(1), while $\psi_K(R)$ is the single-particle
function of the K-electron in residual ion. In our case
$\psi_K(R)$ is just the $1s$ function of the Coulomb field with the charge $Z$.

In the single-particle approximation
C is simply the value of the single-particle wave function at the coordinate origin.
To illustrate the quality of the functions (3) we compare the results for the factor
$C$ calculated by using the CFHH functions and the functions (3).
In the latter case we find an analytic expression
\begin{equation} \label{16}
C=\frac{2 N Z^{3/2}}{\sqrt{\pi}(2Z-1/2)^3}\ ,
\end{equation}
providing $C=0.102$ for the case of atomic helium. The numerical
calculations with the CFHH functions give $C=0.103$ in this case.
Hence, employing the approximate function (3) leads to the error of
$1\%$. Earlier the authors of \cite{16} found that the value of $C$
obtained by using the Hylleraas-type variational function is well
approximated by employing a hydrogenlike function with
$Z_{eff}=Z-0.53$.

Now let us turn to the case of the double photoionization.
The shape of the spectrum curve of the double photoionization changes
with the photon energy growth. The mechanisms which cause these changes
are explained in \cite{3}. While the photon energy $\omega$ is smaller
than certain value $\omega_1$, the energy distribution approaches its
minimum at the central point, with the equal energies of the outgoing
electrons, $\varepsilon_{1,2}$, {\em i.e.}
$\varepsilon_1=\varepsilon_2$. There is a peak at the central point at
$\omega>\omega_1$, which splits into two at $\omega>\omega_2$. Thus,
there is a local minimum at $\varepsilon_1=\varepsilon_2$ at
$\omega>\omega_2$.

The values of $\omega_{1,2}$ were obtained in \cite{17} by using the
CFHH functions. We shall not repeat derivation of the corresponding
equations here. We rather explain their origin and put them down, in
order to illustrate, how the functions (3) enable to obtain
approximate solutions.

The values of $\omega_{1,2}$ can be presented as solutions of the
following equation, which involves the functions $F(R)$ and
$\Phi(R)$ \cite{17}:
\begin{equation} \label{17}
\lambda\mu\ =\ \omega^{9/2}A(\omega)\ ,
\end{equation}
with $\lambda$ being a certain numerical coefficient, and
\begin{equation} \label{18}
\mu\ =\ \int\limits^\infty_0 dr|F(r)|^2\ ,
\end{equation}
while the function $A$ depends on $\omega$ in a more complicated way:
\begin{equation} \label{19}
A(\omega)\ =\ \int\limits^{+1}_{-1}dtt^2(1-2t^2)D(\omega^2t^2)
\end{equation}
with
\begin{equation} \label{20}
D(q^2)=|\int\limits^\infty_0 \frac{\sin(qr)}{qr}\,\Phi(r)r^2dr|^2\ .
\end{equation}

Employing the exact CFHH functions requires tedious computations.
However, using the approximate wave functions (3) one can obtain
analytical expressions for both left-hand side and right-hand side of
Eq.(17). Putting $F(r)=F_A(r)$ and $\Phi(r)=\Phi_A(r)$ we obtain
$\mu=\frac1{2Z-1}$, while
\begin{equation} \label{21}
A(\omega)=\frac1{\omega^6}\left(\frac{6a^6+13a^4+2a^2+3}{6a^2(a^2+1)^3}
+\frac{1-2a^2}{2a^3}\arctan\frac1a\right)
\end{equation}
with $a=2Z/\omega$.

The values of $\omega_{1,2}$ obtained by using the CFHH functions and
the functions (3) are presented in Table 3. One can see that the
discrepancy between two sets of results drops rapidly with $Z$ growth.
Being 22\% for H$^-$ and 9\% for helium, it becomes 4\% for $Z=4$.

\section{Summary}

We build very simple analytical presentations (3),
for the wave functions $F(R)$ and $\Phi(R)$
describing  ground states of two-electron systems
bound by the Coulomb field of the nucleus
in the space regions $r_{1,2}=0$ and $r_{12}=0$.
The presentation is based on the behavior of the
exact solution of the Schr\"odinger equation near the three-particle
coalescence singularity. Comparing our functions (3) to the
locally correct CFHH functions for the ion $H^-$, atomic helium
and light heliumlike ions (relativistic corrections, which are of the order
$(Z/137)^2$ are not included), we found  good agreement in a large
interval of the values of the $R$.
As is evident the precision of the
approximate functions increases with the nuclear charge $Z$ growth.

We show that the solutions of the single-particle Schr\"odinger
equations in the Coulomb fields with the charges $Z_1=Z-1/2$ and
$Z_2=2Z$ provide natural basis for expansion of the functions
$F(R)$ and $\Phi(R)$ with the domination of $1s$ terms. The latter
tendency increases with $Z$. The approach is more precise for the
function $F(R)$, then for $\Phi(R)$.

Examples, presented in Sec.4 show that even for the lightest heliumlike
systems such as $H^-$ and $He$ the wave functions (3) can be used at
least for the estimation of the physical parameters.

The high precision of such a simple approximation that properly treats
singularities in the wave function is in agreement with the
conventional believe that the singularities determine such important
atomic characteristics as high-energy photoionization cross sections.

\begin{acknowledgments}
M.Ya.A. is grateful to the Binational Science Foundation (grant 2002064)
and to the Israeli Science Foundation (grant 174/03)
for financial support of this research. E.G.D. is grateful for the
hospitality extended during his visit to the Hebrew University. The
research of V.B.M. was supported by the Israeli Science Foundation (grant 131/00).
\end{acknowledgments}

\newpage
\begin{table}
\caption{The value $F(0)=\Phi(0)=N$ for several values of $Z$. The
ratio $\tilde r=N/N_c$ with $N_c$ defined by Eq.~(10) illustrates the
convergence to the high $Z$ limit.}

\begin{center}
\begin{tabular}{|c|c|c|c|c|c|c|}

\hline
$Z$ & 1 & 2 & 3 & 4 & 5 & 6\\

\hline

$N$ & 0.071 & 1.37 & 5.77 & 15.2 & 31.6 & 56.8\\

$\tilde r=\frac N{N_c}$ & 0.22 & 0.61 & 0.67 & 0.74 & 0.79 & 0.83\\

\hline
\end{tabular}
\end{center}
\end{table}

\begin{table}
\caption{The coefficients of the two lowest terms of expansions (11) of
the CFHH functions in terms of the Coulomb functions. The coefficients
of the next terms are limited by the conditions $|a_i|<\tilde a$,
$|b_i|<\tilde b$, while the values of $\tilde
a=(1-a^2_{1s}-a^2_{2s})^{1/2}$ and $\tilde b=(1-b^2_{1s}-
b^2_{2s})^{1/2}$ are presented in the two bottom lines.}

\begin{center}
\begin{tabular}{|c|c|c|c|c|} \hline

$Z$ &1 & 2 & 3 & 4\\
\hline

$a_{1s}$ & 0.98482 & 0.99970 & 0.99991 & 0.99996\\

$b_{1s}$ & 0.99067 & 0.99807 & 0.99918 & 0.99955 \\

$a_{2s}$ & --0.144 & --0.020 & --0.010 & --0.007\\

$b_{2s}$ & --0.108 & --0.046 & --0.030 & --0.022\\

$\tilde a$ & 0.097 & 0.015 & 0.008 & 0.005\\

$\tilde b$ & 0.082 & 0.041 & 0.028 & 0.021\\
\hline
\end{tabular}
\end{center}
\end{table}

\begin{table}
\caption{The values of $\omega_1$ and $\omega_2$ (Sect.4) in keV for
the ground states of the lightest heliumlike systems, calculated by
using the CFHH functions \cite{10} and the functions (3).}

\begin{center}
\begin{tabular}{|c|c|c|c|c|} \hline

$Z$ & 1 & 2 & 3 & 4 \\
\hline
$\omega_1$ -- this work & 0.67 & 2.11 & 3.92 & 6.14\\

$\omega_1$ -- \cite{10}  & 0.55 & 1.93 & 3.70 & 5.89\\

$\omega_2$ -- this work  & 4.86 & 9.71 & 14.5 & 19.3\\

$\omega_2$ -- \cite{10}  & 3.97 & 8.89 & 13.7 & 18.5\\
\hline
\end{tabular}
\end{center}
\end{table}

\begin{figure}[p]
\begin{center}
\begin{tabular}{cc}
\epsfxsize=5cm\epsfbox{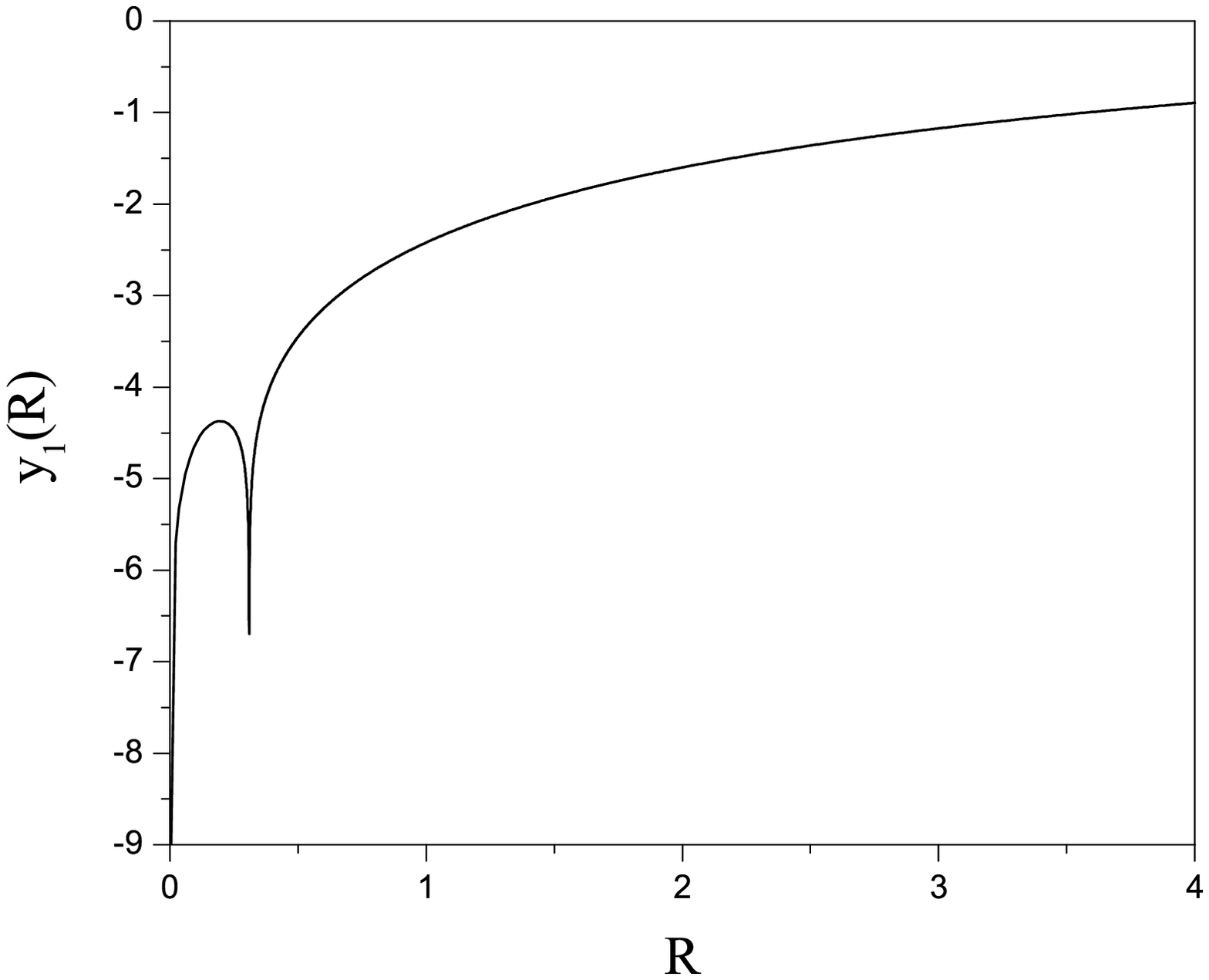}&\epsfxsize=5cm\epsfbox{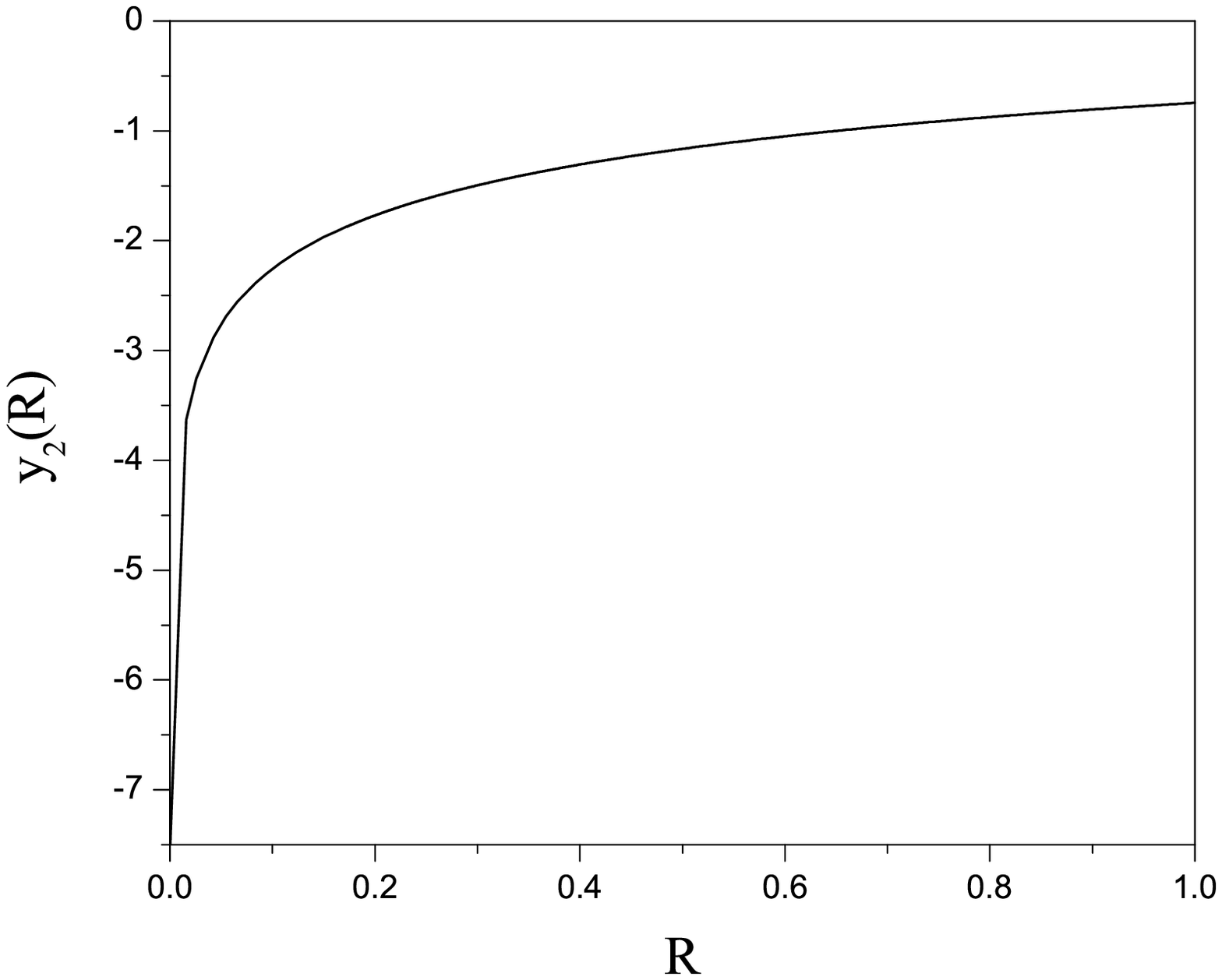}\\
\multicolumn{2}{c}{\bf(a)}\\ &\\
\epsfxsize=5cm\epsfbox{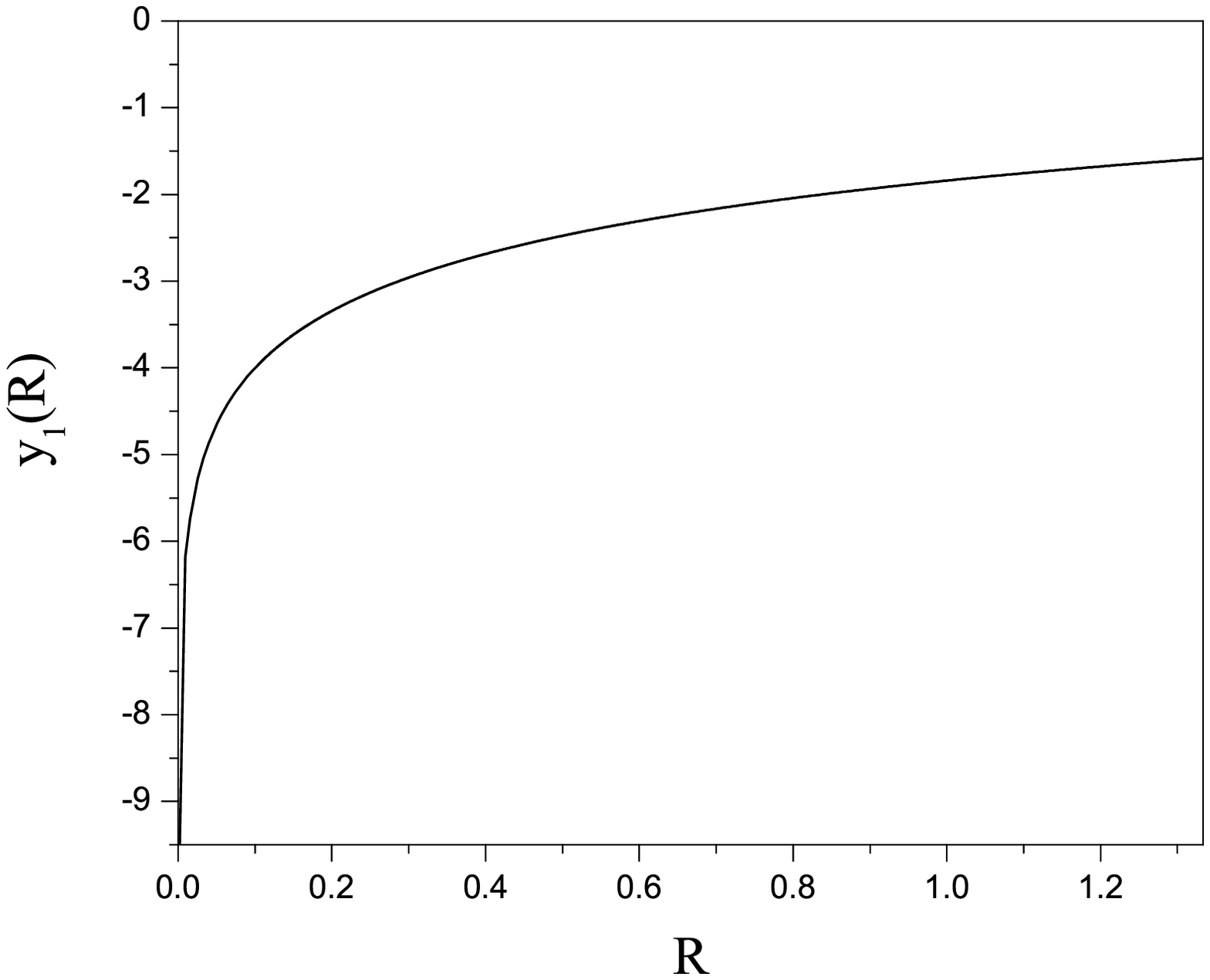}&\epsfxsize=5cm\epsfbox{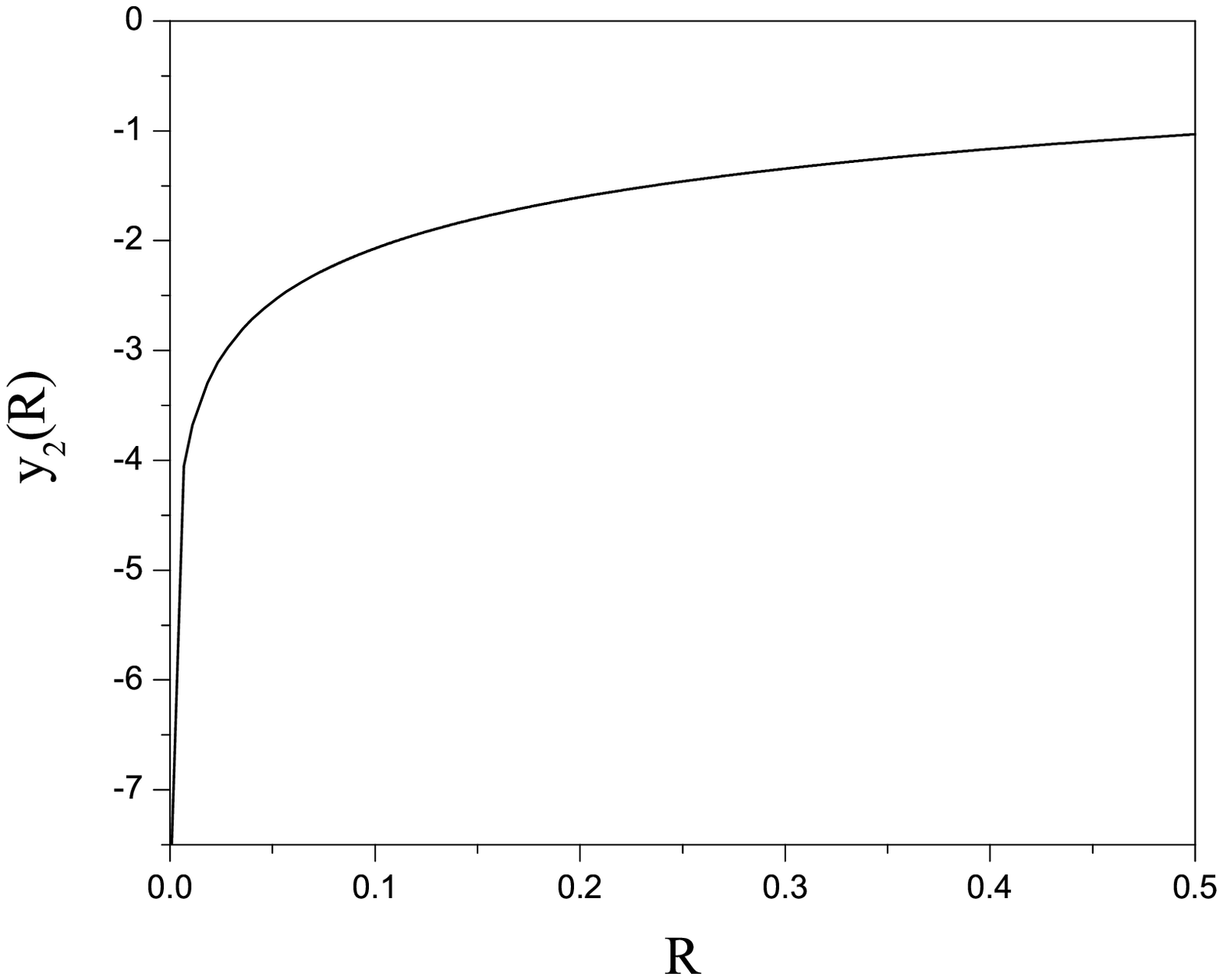}\\
\multicolumn{2}{c}{\bf(b)}\\ &\\
\epsfxsize=5cm\epsfbox{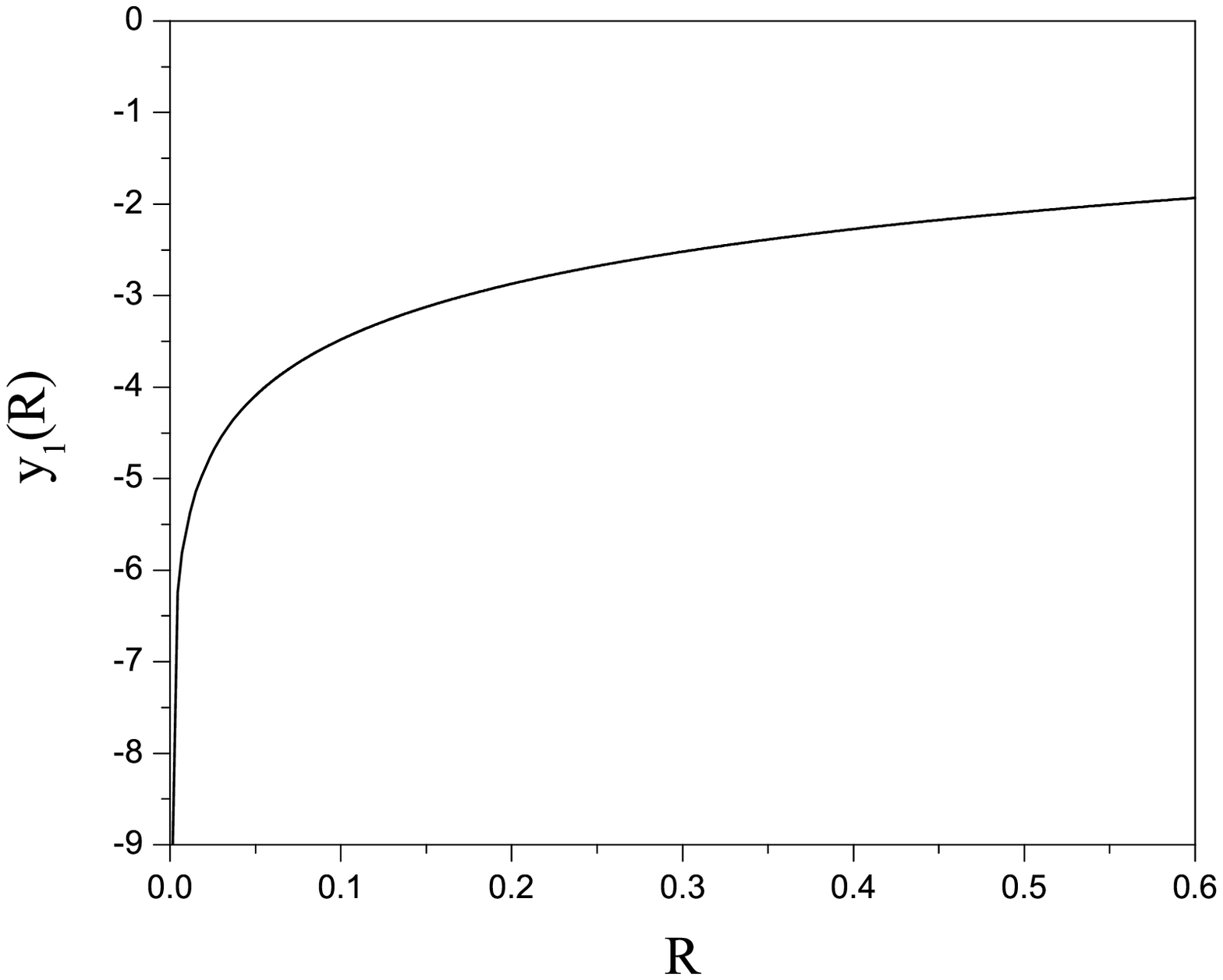}&\epsfxsize=5cm\epsfbox{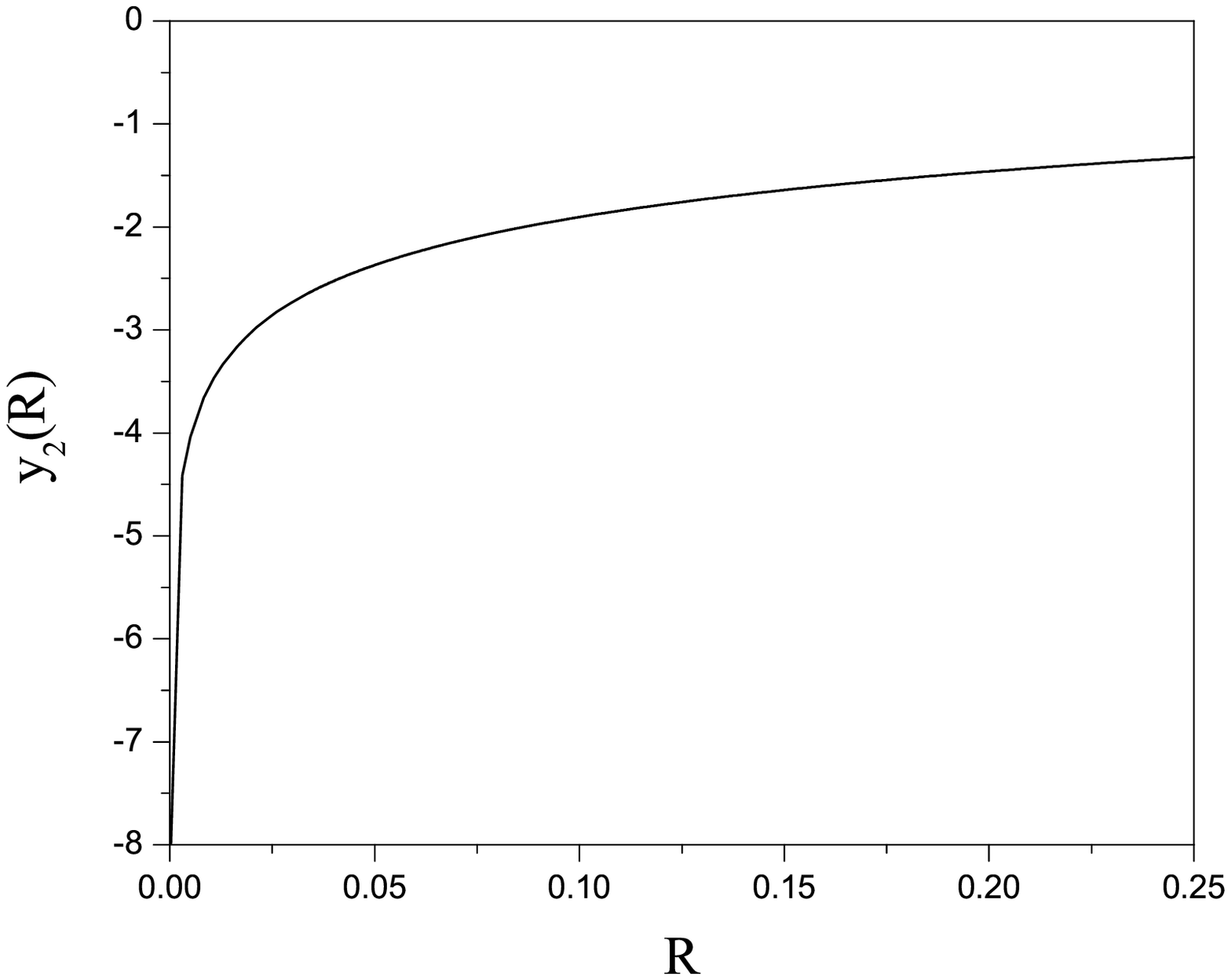}\\
\multicolumn{2}{c}{\bf(c)}
\end{tabular}
\end{center}
\caption{ The functions $y_1(R)$ and $y_2(R)$ defined by Eq. (9),
for the negative ion H$^-(Z=1)$ \newline --- Fig.1a, for atomic
helium $(Z=2)$
--- Fig.1b, and for the ion Be$^{++}(Z=4)$ --- Fig.1c.}
\end{figure}
\clearpage

\begin{figure}
\begin{center}
\epsfig{file=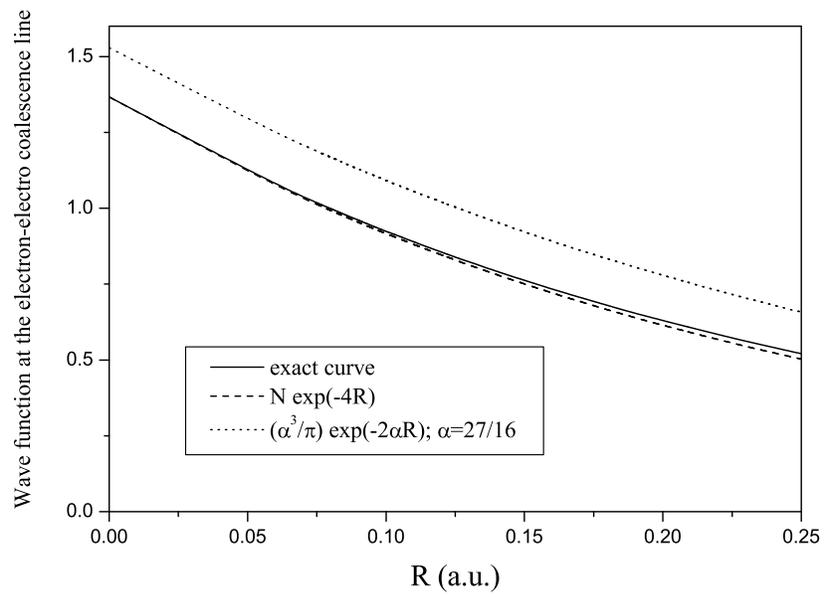,width=120mm}
\end{center}
\caption{The exact and the approximate helium wave functions at the
electron-electron coalescence line.} \label{fig2}
\end{figure}

\end{document}